\documentclass[12pt]{iopart}

\usepackage{graphicx}
\usepackage{colortbl}
\usepackage{iopams}
\usepackage{amssymb}
\usepackage{bbm}

\newcommand{\bolds}{\mathbf{s}}

\begin{document}
\title{Universal Quantum Computing with Spin and Valley}
\author{Niklas Rohling and Guido Burkard}
\address{Department of Physics, University of Konstanz, D-78457 Konstanz, Germany}
\date{\today}

\begin{abstract}
We investigate a two-electron double quantum dot with both spin and valley
degrees of freedom as they occur in graphene, carbon nanotubes, or silicon,
and regard the 16-dimensional space with one electron per dot as a
four-qubit logic space.
In the spin-only case, it is well known that the exchange coupling
between the dots combined with arbitrary single-qubit operations is 
sufficient for universal quantum computation.
The presence of the valley degeneracy in the electronic band structure alters the form of the
exchange coupling and in general leads to spin-valley entanglement.
Here, we show that universal quantum computation 
can still be performed by exchange interaction and single-qubit gates in
the presence of the additional (valley) degree of freedom.
We present an explicit pulse sequence for a spin-only controlled-NOT consisting of 
the generalized exchange coupling and single-electron spin and valley rotations.
We also propose state preparations and projective measurements 
with the use of adiabatic transitions between states with (1,1) and
(0,2) charge distributions similar to the spin-only case, but with the 
additional requirement of controlling the spin and the valley Zeeman
energies by an external magnetic field.
Finally, we demonstrate a universal two-qubit gate between a spin and 
a valley qubit, allowing universal gate operations on the combined
spin and valley quantum register.
\end{abstract}

\maketitle


\section{Introduction}
Since Loss and DiVincenzo \cite{LoDi1998} proposed quantum computing with electron spins
in double quantum dots, there has been a substantial experimental progress in the field of
coherent spin manipulation in semiconductors \cite{petta,koppens,foletti2009, brunner,hanson_review}.
The majority of these experiments has been performed in gallium
arsenide (GaAs) where the electron spin suffers from decoherence
due to its coupling to a typically  large number of nuclear spins, as
well as spin relaxation due to spin-orbit coupling.

In carbon materials such as graphene or carbon nanotubes (CNTs), the hyperfine interaction is
much weaker because $^{13}$C is the only naturally occurring carbon isotope carrying a nuclear spin
and the amount of $^{13}$C in natural carbon is merely $\sim 1 \%$.
Similar considerations hold for quantum dots based on silicon (Si) and
germanium (Ge), where less
than 5 \%  (8 \%)  of all naturally occurring Si (Ge) atoms carry a nuclear spin.
In graphene, the spin-orbit coupling is also expected to be weak \cite{recher_trauzettel}.

However, the situation for quantum dots in graphene and CNTs compared to GaAs
is complicated by the presence of an additional orbital degree of freedom, the so-called valley
iso-spin \cite{trauzettel,recher_trauzettel}, with basis states $|K\rangle$ and $|K'\rangle$,
denoting the two inequivalent Dirac points in the first Brioullin zone
in the graphene band structure.
Experimentally, spin states in graphene quantum dots have been identified by transport
measurements \cite{PhysRevLett.105.116801} but valley states have not
been observed yet,
whereas in CNTs, a fourfold grouping of electronic states due to spin and valley degree
of freedom have already been observed for a decade in transport
measurements \cite{Buitelaar2002,liang,cobden}.
The relaxation and dephasing times of two valley- and spin-degenerate
electrons in a CNT double quantum dot have been studied
experimentally \cite{churchill} by using both transport measurements
in the Pauli blockade regime \cite{palyi2010}, as well as pulsed-gate
measurements \cite{reynoso,reynoso2}.

Interestingly, the situation for quantum dots in Si/SiGe heterostructures is similar
since the six-fold valley degeneracy in bulk silicon is partially lifted
in strained systems \cite{schaeffler}, giving rise to a remaining
two-fold valley degeneracy.
The confining potential can lead to a further splitting of the remaining two valley states,
which ultimately leads back to spin-only qubits and operations \cite{culcer}.
In recent experiments with silicon-based quantum dots, coherent spin manipulation
with the exchange interaction has been performed successfully \cite{maue},
and some control over the valley splitting has been demonstrated \cite{goswami}.
Both in Si \cite{eriksson} and in graphene \cite{rycerz,trauzettel,PhysRevB.76.235404}
there have been speculations that the valley degree of freedom
might serve as an additional resource for classical or quantum
information processing, i.e. as a classical bit for valleytronics
\cite{rycerz,trauzettel} or as a qubit \cite{eriksson,PhysRevB.76.235404}.
However, the presence of an orbital (e.g., valley) degeneracy 
leads to the following difficulty for quantum computing.
The additional degree of freedom modifies the form of the 
exchange interaction which is based on the Pauli exclusion principle.
E. g., a spin triplet in the (1,1) charge configuration may not be
blocked from tunneling to a (0,2) state if the two electrons reside in
different valleys.  Here, $(m,n)$ stands for $m$ electrons in the left
and $n$ electrons in the right quantum dot.
Such a valley-dependent spin-exchange leads to spin-valley
entanglement and implies that the controlled-NOT (CNOT) gate cannot 
be performed in the same way as proposed in \cite{LoDi1998} as long as
the valley degeneracy is present \cite{trauzettel}.
\begin{figure}[t]
\begin{center}
\includegraphics[clip,width=0.8\textwidth]{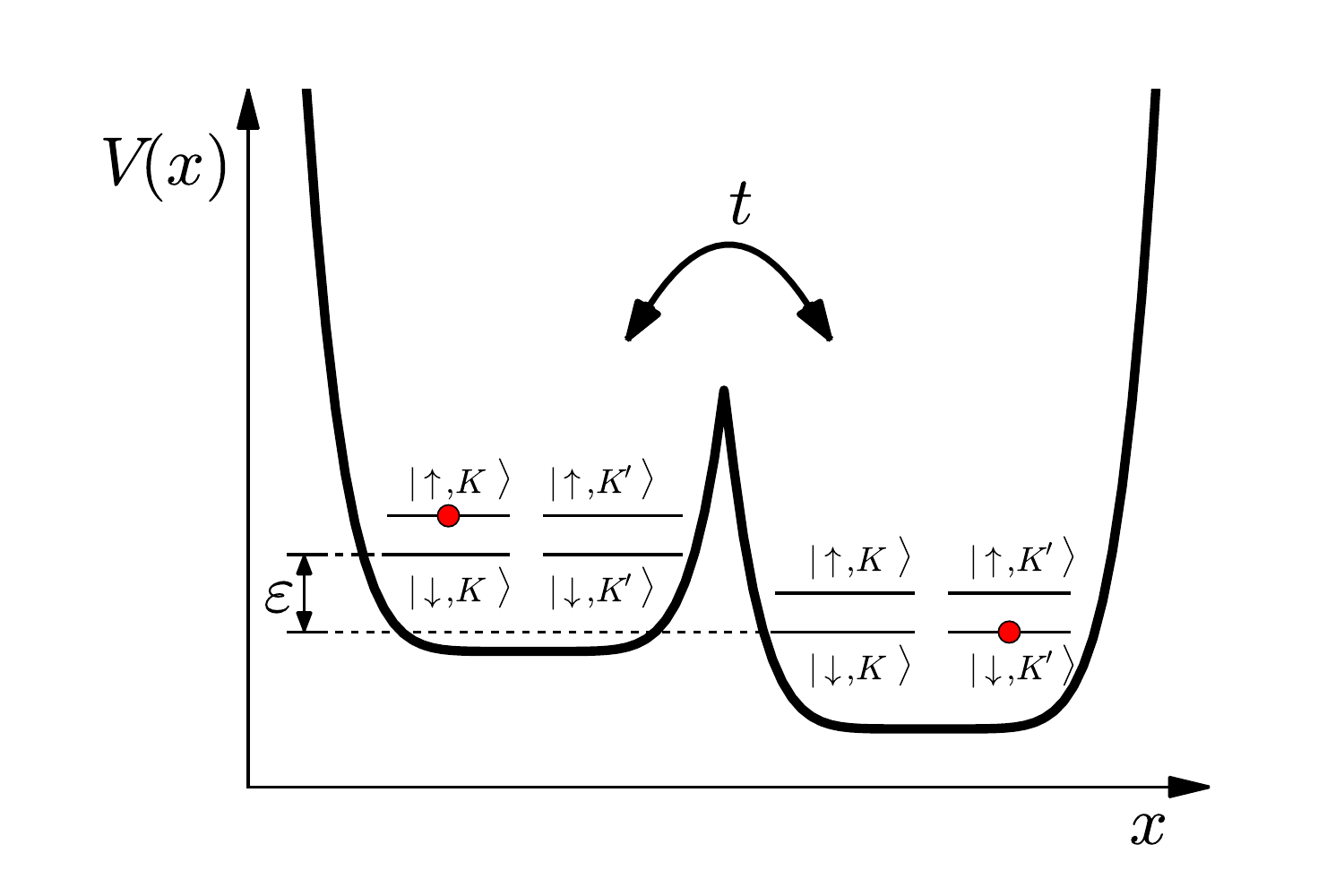}
\caption{Schematic of a double quantum dot formed by a confinement
  potential $V(x)$ and filled with two electrons (red dots). 
In the presence of valley and spin
degeneracy there are 16 states with one electron in each dot, i.e. in
the (1,1) charge configuration. 
In the example shown here, the two-electron state is
$|s_1,s_2,\tau_1,\tau_2\rangle=|\uparrow,\downarrow,K,K'\rangle$.
The hopping (tunneling) matrix element between the dots and the 
inter-dot bias energy are denoted by $t$ and $\varepsilon$.
\label{fig:sketch}}
\end{center}
\end{figure}

Therefore, proposals for graphene quantum dots have attempted to avoid the valley degeneracy
\cite{recher_trauzettel} by using armchair boundary condition for quantum dots in a
graphene nanoribbon \cite{trauzettel} or by applying a magnetic field perpendicular to
the graphene sheet for quantum dots defined by electrostatic gates \cite{PhysRevB.79.085407}.
In a recent proposal, Wu \textit{et al.} suggest to use only the
valley degree of freedom as a qubit and fix
the spin degree of freedom  by a strong in-plane magnetic field \cite{PhysRevB.84.195463}.

In this paper, we consider a double quantum dot with two electrons and
regard both spin and valley degrees of freedom as potential qubits.
This leads to a 16-dimensional logic space consisting of two spin and two valley qubits
(see Fig. \ref{fig:sketch}).
We show that it is possible to perform a CNOT gate as a universal
two-qubit gate exclusively
on the spin or the valley qubits if the exchange interaction and single-qubit
manipulations can be implemented.
For singlet-triplet qubits the exchange interaction directly produces
a CNOT gate, up to single-qubit operations.
Furthermore, we investigate how state preparation and measurements can
be carried out by adiabatically changing the asymmetry between the
dots with the use of the appropriate gate voltage control.
An external magnetic field turns out to be important for both
preparation and measurement.
The field allows one to break the six-fold degeneracy of the states
with both electrons in the same dot, (2,0) and (0,2), and thus allows for the selective
preparation of one such state in the initialization process.
The magnetic field also selects the states that are driven from a
symmetric (1,1) back to this
asymmetric (2,0) or (0,2) charge state.
For quantum state read-out, the resulting charge state can then be measured with a charge
detector, e.g., a nearby quantum point contact \cite{hanson_review}.
We explain below how a projective measurement on one specific state can be achieved by three
charge measurements under different configurations of the magnetic
field or, alternatively, with a constant magnetic field and the help of single-qubit operations.

This paper is organized as follows. In Sec.~\ref{sec:model}, we introduce the model Hamiltonian
for the tunnel-coupled double quantum dot with two electrons, and
derive the general form of the exchange
interaction without a magnetic field (Sec.~\ref{sec:xi}), and
including a magnetic field (Sec.~\ref{sec:mf}).
Sec.~\ref{sec:cnot} contains a pulse sequence for the CNOT gate.
Our considerations and results concerning state preparation and measurement are 
reported in Sec.~\ref{sec:statprep_meas}.
In Sec. \ref{sec:qregister}, we describe how a quantum register using spin and valley qubits
may be constructed by using singlet-triplet qubits in two quantum dots
and usual single-electron spin and valley qubits in the other dots.
Conclusions are drawn and an outlook towards possible further investigations is
given in Sec.~\ref{sec:concs}.


\section{Model}
\label{sec:model}
We consider two electrons in a double quantum dot described by the Hamiltonian
\begin{equation}
\label{eqn:hamiltonian}
 H = H_0+H_T+H_B,
\end{equation}
where the two quantum dots with one orbital each are described by
\begin{equation}
\label{eqn:hamiltonian1}
 H_0 =   \frac{\varepsilon}{2}(\hat n_1-\hat n_2) + U\sum_{j=1,2}\hat n_j(\hat n_j-1),
\end{equation}
with $\varepsilon$ denoting the difference between the energy levels of the two dots,
controllable by gate voltages (Fig.~\ref{fig:sketch}).
The additional Coulomb energy of two electrons in the same dot is
denoted by $U$.
The number operators $\hat{n}_j$ ($j=1,2$) include a sum over the spin 
$s=\uparrow,\downarrow$ and the valley degree of freedom 
$\tau =\pm\equiv K,K'$, 
\begin{equation}
 \hat n_j=\sum_{s,\tau}\hat c^\dagger_{js\tau}\hat c_{js\tau},
\end{equation}
where $\hat{c}_{j,s\tau}^{\left(\dagger\right)}$ annihilates (creates) an electron in
the $j$th quantum dot with spin and valley quantum numbers $s$ and $\tau$.
In the spin-only case, the Hilbert space for this model of a double
quantum dot consists of four states with a $(1,1)$ charge distribution, one $(0,2)$
and one $(2,0)$ charge state \cite{LoDi1998}, where $(n,m)$ denotes a
state with $n$ electrons in the left and $m$ electrons in the right dot. 
No further states with two electrons in one dot with a single orbital are
permitted by the Pauli principle.
Including the valley degree of freedom, we end up with 16 $(1,1)$ states, six $(0,2)$ states,
and six $(2,0)$ states.

\subsection{Exchange interaction}
\label{sec:xi}
The two quantum dots are coupled by the spin- and valley-preserving
hopping (tunneling), 
\begin{equation}
  H_T = t\sum_{s\tau} \left(\hat c^\dagger_{2,s\tau}\hat c_{1,s\tau}+h.c.\right),
\end{equation}
where $t$ denotes the tunneling matrix element.
We first consider the case without a magnetic field, $H_B=0$, 
and the parameters in the regime $|t|\ll|U\pm\varepsilon|$
where the (1,1) charge states are approximate
eigenstates of the Hamiltonian (\ref{eqn:hamiltonian}).
The Pauli principle implies that only those (1,1) states are coupled to (0,2)
and (2,0) states which are antisymmetric in the combined spin and valley space.
In spin space, there is one antisymmetric state for two electrons, the
spin singlet, and there are three symmetric states, the spin triplet states; 
for the valley space alone, the situation is analogous.
To study the symmetric and antisymmetric states in the combined
spin and valley space, we introduce vectors of Pauli matrices for 
the spin and valley of the electron in the first ($j{=}1$) or second
($j{=}2$) quantum dot, as
$\bolds_j=(s_{jx},s_{jy},s_{jz})^T$ and $\btau_j=(\tau_{jx},\tau_{jy},\tau_{jz})^T$,
and express the projection on the singlet (upper index $S$) and the triplet
(upper index $T$) sector as follows,
\begin{eqnarray}
P_{\rm spin}^S &=& \frac{1-\bolds_1\cdot\bolds_2}{4},
       \hspace{.5cm} P_{\rm spin}^T = \frac{\bolds_1\cdot\bolds_2 + 3}{4},\\
 P_{\rm valley}^S &=& \frac{1-\btau_1\cdot\btau_2}{4},
       \hspace{.5cm} P_{\rm valley}^T = \frac{\btau_1\cdot\btau_2 + 3}{4}.
\end{eqnarray}
These operators fulfill the usual relation for projectors,
$\left(P_{F}^q\right)^2=P_F^q$ and $P_{F}^S+P_{F}^T=1$, 
where $F={\rm spin},{\rm valley}$ and $q=S,T$.
The projection operator on the antisymmetric states of the combined
spin and valley space is given by $P_{\rm as} = P_{\rm spin}^S P_{\rm valley}^T + P_{\rm spin}^T
P_{\rm valley}^S$ and defines the effective low-energy Hamiltonian for the (1,1) states,
\begin{equation}
\label{eqn:Heff}
 H_{\rm eff} = -JP_{\rm as}
             =  \frac{J}{8}\big( (\bolds_1\cdot\bolds_2)(\btau_1\cdot\btau_2)
                + \bolds_1\cdot\bolds_2 + \btau_1\cdot\btau_2 - 3\big).
\end{equation}
The exchange coupling $J$ is given by $J = 4 t^2U/(U^2-\varepsilon^2)$, which can be
determined by a Schrieffer-Wolff transformation on $H$ in the same way
as it is used in the spin-only case \cite{burkard_imamoglu}, see \ref{sec:appendix}.
The eigenvalues of $H_{\rm eff}$ are $-J$ and 0 with a six- and a ten-dimensional
eigenspace, respectively (see also \cite{reynoso}).
The projection on the subspace with eigenenergy $-J$ is given in terms of spin and valley
operators but for the exchange coupling the origin of the degeneracy is irrelevant.
Hence, the result we obtained here is true for any four-fold degeneracy of the electron,
provided that tunneling conserves this four-valued internal quantum number \cite{choi}.

We can consider the reduced Hilbert space of the (1,1) states
belonging to $H_{\rm eff}$ as a four-qubit space with the
spins in the first and in the second quantum dot as the first and second qubit and
the valley iso-spins as the qubits number three and four, with $|\uparrow\rangle\equiv|0\rangle$,
$|\downarrow\rangle\equiv|1\rangle$, $|+\rangle\equiv|0\rangle$, $|-\rangle\equiv|1\rangle$.
Using the four Bell states,
\begin{equation}
\label{eqn:bell}
 |\phi_\pm\rangle = \frac{|00\rangle\pm|11\rangle}{\sqrt{2}}, \hspace{.5cm}
 |\psi_\pm\rangle = \frac{|01\rangle\pm|10\rangle}{\sqrt{2}},
\end{equation}
as basis states in spin and valley space, and building a product basis,
\begin{equation}
\{|\phi_+\rangle,|\phi_-\rangle,|\psi_+\rangle,|\psi_-\rangle\}_{\rm spin}\otimes
\{|\phi_+\rangle,|\phi_-\rangle,|\psi_+\rangle,|\psi_-\rangle\}_{\rm  valley} ,
\label{eqn:doublebell}
\end{equation}
the corresponding matrix of $H_{\rm eff}$ becomes diagonal.
Obviously, we can identify $|\psi_-\rangle$ with the singlet and the other three vectors
with the triplet space of the spin or the valley.  We call Eq.~(\ref{eqn:doublebell}) 
the \textit{double Bell basis}.

\subsection{Magnetic field}
\label{sec:mf}
The influence of a magnetic field on the spin and valley is given by 
\begin{equation}
 H_B = \sum_{j=1,2}h_{Sj}(\hat n_{j\uparrow}- \hat n_{j\downarrow})
      + \sum_{j=1,2}h_{Vj}(\hat n_{j+}- \hat n_{j-}),
\end{equation}
where the number operators are defined as 
$\hat n_{js} = \sum_{\tau}\hat c^\dagger_{js\tau}\hat c_{js\tau}$ 
and $\hat n_{j\tau} = \sum_{s}\hat c^\dagger_{js\tau}\hat c_{js\tau}$.
The parameter $h_{Sj}$ denotes the spin Zeeman energy in the $j$th
quantum dot, where the spin quantization axis is chosen along the
direction of the magnetic field.
The valley degeneracy in each dot is broken by the magnetic-field component parallel to the axis
of a CNT or orthogonal to the graphene sheet. This splitting is expressed by $h_{Vj}$ which we refer
to as the valley Zeeman energy.
It has been shown experimentally for a CNT \cite{kuemmeth} and theoretically for graphene quantum
dots \cite{PhysRevB.79.085407} that the valley Zeeman splitting due to this component of the
magnetic field is much larger than the corresponding spin Zeeman splitting.
On the other hand, the in-plane components in graphene and the components orthogonal to the
axis of a CNT mainly influence the spin Zeeman energy. Therefore, the
values of $h_{Sj}$ and $h_{Vj}$ can be set nearly independently by an external magnetic field.

We neglect here that the magnetic fields in the dots can have different directions, which would lead
to additional avoided crossings in the spectrum of $H$.
Under this condition, we still can apply the Schrieffer-Wolff transformation used in
\cite{burkard_imamoglu} to obtain an effective Hamiltonian for the 16 (1,1) states, see \ref{sec:appendix}.
We define
$h_V =(h_{V1}+h_{V2})/2$, $h_S =(h_{S1}+h_{S2})/2$, $\Delta h_V
=h_{V1}-h_{V2}$, and $\Delta h_S=h_{S1}-h_{S2}$.
In the limit $|t|, |\Delta h_V|, |\Delta h_S|\ll|U\pm\varepsilon|$ we find
\begin{eqnarray}
\label{eqn:HeffB}
H_{{\rm eff},B} &=& \frac{J}{8}\big((\bolds_1\cdot\bolds_2)(\btau_1\cdot\btau_2)
                  + \bolds_1\cdot\bolds_2 \! +\!  \btau_1\cdot\btau_2 - 3\big)
                +\!\!\sum_{j=1,2}(h_{Sj}s_{jz}+h_{Vj}\tau_{jz}) \nonumber\\
                  & & +\frac{J\varepsilon}{U^2 \!-\! \varepsilon^2}
  \! \left[\frac{\btau_1 \!\cdot\! \btau_2  \!-\! 1}{4}(s_{1z}\!+\!s_{2z})\Delta h_S 
+\frac{\bolds_1\!\cdot\!\bolds_2 \!-\! 1}{4}(\tau_{1z} \!+\! \tau_{2z})\Delta h_V\right]. \nonumber\\
\end{eqnarray}
The magnetic field might be a resource for tuning the exchange interaction, particularly in situations
where the gradient is large and the linear approximation given here
is not valid (for a more general expression, see \ref{sec:appendix}).
Nevertheless, we consider quantum gates created by the exchange coupling without a
magnetic field in the following. More precisely, we assume that
$\Delta h_S$ and $\Delta h_V$ are negligible while the exchange
coupling is applied.  This can be achieved if $J$ as a function of time is tuned by varying $\varepsilon$.
The parts of the Hamiltonian $H_{{\rm eff},B}$ which depend on $h_S$ and $h_V$ commute with $H_{\rm eff}$
and can therefore be regarded as single-qubit operations performed before or after the
exchange coupling is applied.


\section{CNOT gate on spin qubits}
\label{sec:cnot}
In this section, we show that it is possible to perform a
controlled-NOT (CNOT) gate on the spin qubits alone,
\begin{equation}
 \mathrm{CNOT}_{\rm spin}  =  \mathrm{CNOT}\otimes\mathbbm{1}
= \left(\begin{array}{cccc} 1 & 0 & 0 & 0\\
                            0 & 1 & 0 & 0\\
                            0 & 0 & 0 & 1\\
                            0 & 0 & 1 & 0
        \end{array}\right)
\otimes
\left(\begin{array}{cccc} 1 & 0 & 0 & 0\\
                            0 & 1 & 0 & 0\\
                            0 & 0 & 1 & 0\\
                            0 & 0 & 0 & 1
        \end{array}\right),
\label{eqn:spinCNOT}
\end{equation}
by applying the exchange interaction Eq.~(\ref{eqn:Heff}), supplemented with single-qubit
operations on both the spin and valley qubits.
Note that in Eq.~(\ref{eqn:spinCNOT}), the matrices are represented in
the product basis of the qubit states (not in the Bell basis).
Because CNOT gates can be combined with single-qubit gates to
form arbitrary unitaries on any number of qubits
\cite{divincenzo1995,barenco1995}, our result below implies that universal quantum computing
in the subspace of the spin qubits can be realized with the exchange
interaction and single-qubit gates, despite the presence of the valley degeneracy.
For an explicit construction of a CNOT gate, 
we define the time-evolution operator $U(\phi)$ of the exchange interaction as
\begin{equation}
 U(\phi) = e^{-i\int_0^{t_e}dt'\, H_{\rm eff}(t')} = \mathbbm{1} +
 \left(e^{i\phi}-1\right)P_{\rm as} ,
\end{equation}
where $\phi=\int_0^{t_e}dt'\,J(t')$ is the time-integrated exchange coupling and $H_{\rm eff}$ is the exchange
Hamiltonian defined in Eq.~(\ref{eqn:Heff}).
In the absence of the valley degeneracy, e.g., $\tau_1=\tau_2=K$ and thus $\btau_1\cdot\btau_2=1$ in
Eq.~(\ref{eqn:Heff}), the exchange interaction
directly generates a $\sqrt{\mathrm{SWAP}}$ gate for $\phi=\pi/2$,
\begin{equation}
\sqrt{\mathrm{SWAP}} = \frac{1+i}{2} \mathbbm{1} +\frac{1-i}{2} \mathrm{SWAP}, 
\end{equation}
for the spin qubits, which can be applied twice in combination with
single-spin rotations to generate CNOT \cite{LoDi1998}.  
Here, the $\mathrm{SWAP}$ gate simply exchanges the states of the two
spin qubits.  While the $\mathrm{SWAP}$ gate itself can also be obtained from the
exchange interaction, it is not sufficient to construct CNOT.

In the presence of the valley degeneracy, a gate that interchanges the
spin and valley qubits independently can be obtained similarly as in \cite{LoDi1998}, 
as $U(\pm\pi)=\mathrm{SWAP}\otimes\mathrm{SWAP}$, or explicitly,
$U(\pm\pi)|s_1,s_2,\tau_1,\tau_2\rangle=|s_2,s_1,\tau_2,\tau_1\rangle$.
However, $U(\pm\pi/2) \neq \sqrt{\rm SWAP}\otimes\sqrt{\rm SWAP}$;
instead, we find
\begin{equation}
U(\pm\pi/2) = \frac{1\pm i}{2} \mathbbm{1} +\frac{1\mp i}{2}  {\rm SWAP} \otimes {\rm SWAP}.
\end{equation} 
In addition to producing the required entanglement between the two
spins (and between the two valley iso-spins), this gate simultaneously
also produces entanglement between spin and valley.
To perform CNOT on  the spin (or valley) alone, we thus need a
modified pulse sequence.  We find the following solution,
\begin{equation}
\label{eqn:sswap}
\sqrt{\rm SWAP}_{\rm spin}\equiv \sqrt{\rm SWAP}\otimes\mathbbm{1} 
= e^{i\frac{3\pi}{4}}\big(U(\pi/4)\tau_{1x}U(\pi/4)\tau_{1z}\big)^2,
\end{equation}
from which we can construct a spin-only CNOT, using the result of Ref.~\cite{LoDi1998},
\begin{equation}
\label{eqn:cnot}
\mathrm{CNOT}_{\rm spin} = i e^{-i\frac{\pi}{4}s_{2y}} e^{i\frac{\pi}{4}(s_{2z}-s_{1z})}
                                    \sqrt{\rm SWAP}_{\rm spin} e^{-i\frac{\pi}{2}s_{1z}} \sqrt{\rm SWAP}_{\rm spin}
                                   e^{i\frac{\pi}{4}s_{2y}},
\end{equation}
where the signs of the spin rotations about the $z$ axis are opposite
if one uses another root of SWAP, given as $\sqrt{\rm SWAP}_{\rm spin}^{-1}$, as in
\cite{LoDi1998}.  
Note that $\sqrt{\rm SWAP}_{\rm spin}^{-1}=\sqrt{\rm SWAP}_{\rm
  spin}^{*}$ can be implemented (up to a phase) by replacing $\pi/4$ by
$-\pi/4$ in Eq.~(\ref{eqn:sswap}).
The spin $y$-rotations in Eq.~(\ref{eqn:cnot}) implement a basis
change that transforms CPHASE to the equivalent CNOT.
The single-qubit gates $\tau_{1x}$ and $\tau_{1z}$ in Eq.~(\ref{eqn:sswap}) on the
first valley qubit are implemented as $\exp(i\frac{\pi}{2}\tau_{1\beta})=i\tau_{1\beta}$
where $\beta=x,y,z$.
Another possibility to write the sequence for $\sqrt{\rm SWAP}_{\rm spin}$ is
\begin{equation}
\label{eqn:sswap2}
\sqrt{\rm SWAP}_{\rm spin}=e^{-i\frac{\pi}{4}} U(\pi/4)
                                    \prod_{\beta=x,y,z}\tau_{1\beta}U(\pi/4)\tau_{1\beta},
\end{equation}
which reflects the symmetry of the gate under permutation of the Pauli matrices
$\tau_{1x}$, $\tau_{1y}$, and $\tau_{1z}$.
Note that Eq.~(\ref{eqn:sswap2}) can easily be checked because $U(\phi)$,
$\tau_{1\beta}U(\pi/4)\tau_{1\beta}$ ($\beta=x,y,z$), 
and $\sqrt{\rm SWAP}_{\rm spin}$ are diagonal in the double Bell basis
(\ref{eqn:doublebell}).
Equation (\ref{eqn:cnot}) describes a CNOT gate for the spin qubits that does not
affect the valley states.
The fact that a CNOT gate exclusively on the spin qubits can be performed as in
Eq.~(\ref{eqn:cnot}) by using the new sequence Eq.~(\ref{eqn:sswap})
in a valley-degenerate system is the first main result of this article.
By simply exchanging the single-qubit spin and valley operators
($s\leftrightarrow \tau$) in the equations above, we also find a CNOT
gate in valley space which does not affect the spins.
Here, we have assumed that arbitrary single-qubit operations in spin
and valley space are available.
The implementation of valley rotations within nanosecond time scales using electron
valley resonance in a CNT has been proposed in
\cite{PhysRevLett.106.086801}.
Finally, we note that full valley coherence is not required by the
``valley-assisted''
spin-qubit gate $\sqrt{\rm SWAP}_{\rm spin}$, Eq.~(\ref{eqn:sswap}), and thus for
CNOT, because the spin and valley operations ultimately factorize.
Even if the initial valley state is mixed, the valley iso-spin will be 
disentangled by the end of the gate operation, leaving the spin
qubit sector  coherent. 
However, there is a somewhat less stringent restriction on valley
coherence:  Any valley qubit error (bit or phase flip) which occurs
\textit{during} the gate operation can propagate into the spin sector.
While we do not have sufficient experimental data on valley coherence
to decide whether this condition will be fulfilled,
we note that at least this condition is much easier to satisfy than
full valley coherence.
Starting from the estimated Rabi period for electron valley resonance
\cite{PhysRevLett.106.086801}, we expect the relevant gate operation
time to be around 10~ns.


\section{State preparation and measurement}
\label{sec:statprep_meas}

Before we describe how state preparation and projective measurements
can be performed in a valley-degenerate
system, we briefly characterize the situation in the spin-only case, which has already been
explored experimentally \cite{petta,brunner,hanson_review}.
In a double quantum dot without a valley degree of freedom, the Pauli
principle allows only one state with a (0,2) charge distribution.
In the case $\varepsilon\gg U$ this is the ground state of the system.
Therefore, state preparation is possible by waiting at a large value
of $\varepsilon$ until the double quantum dot relaxes to this ground state.
Afterwards, $\varepsilon$ can be reduced to zero adiabatically which
drives the system to one specific (1,1)
charge state, selected by the magnetic field
(for $B=0$, the spin singlet).
Reading out a qubit state can be achieved by increasing $\varepsilon$
adiabatically, thus allowing a projective measurement on the one
specific (1,1) state that is connected to the (0,2) state,
while all other states remain in a (1,1) charge distribution.
The charge distribution can then be measured with a charge sensor,
e.g., a quantum point contact.

\begin{figure}[htbp]
\begin{center}
\includegraphics[clip,width=0.75\textwidth]{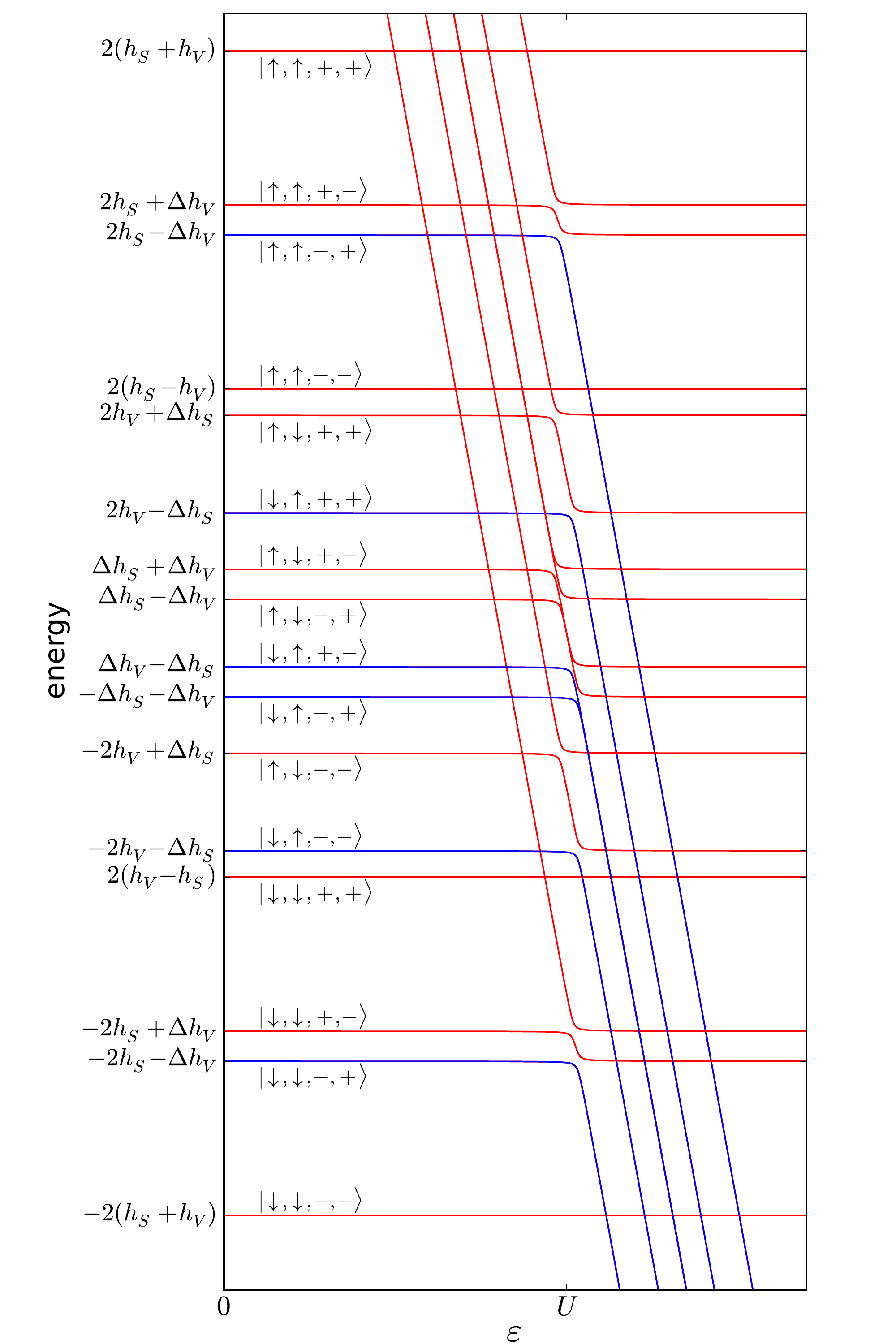}
\caption{Double-dot two-electron energy spectrum described 
by Eq.~(\ref{eqn:hamiltonian}) in dependence of the asymmetry $\varepsilon$.
Here, the magnetic field fulfills $\Delta h_S>\Delta h_V>0$ and the exchange energy at $\varepsilon=0$ is
small compared to $h_F$ and $\Delta h_F$ with $F=S,V$.
The six darker (blue) lines indicates which states are connected to the (0,2) space by an adiabatic
transition, while the brighter (red) lines denote states that remain in the (1,1) space even at
large asymmetries. Note that the center dark (blue) line is two-fold degenerate in the limit of large
$\varepsilon$.}
\label{fig:spectrum}
\end{center}
\end{figure}

In the presence of the valley degree of freedom, the situation is more
complicated because there are six linearly independent (0,2) states.
In order to prepare the system in a well known initial state by a
relaxation process, this sixfold degeneracy has to be lifted.
This can be done using the spin or valley Zeeman term, i.e., by
applying a magnetic field.
Measuring the charge state after increasing the value of $\varepsilon$
realizes a projection on a six- or a ten-dimensional subspace, when the system
goes over to a (0,2) charge state or stays in a (1,1) state, respectively.
To achieve a projective measurement on a single quantum state, several 
charge measurements can be performed in series.
By applying a proper external magnetic field it is possible to influence
which states are connected to a (0,2) state by the adiabatic transition
described above.
Assuming that for $\varepsilon=0$ the exchange interaction $J=4t^2/U$ is small compared to $h_F$ and
$\Delta h_F$ with $F=S,V$, the states $|s_1,s_2,\tau_1,\tau_2\rangle$ with $s_j=\uparrow,\downarrow$
and $\tau_j=\pm$ are approximate eigenstates of the Hamiltonian (\ref{eqn:hamiltonian}).
Fig.~\ref{fig:spectrum} shows the eigenenergies as a function of $\varepsilon$ for the situation
$\Delta h_S>\Delta h_V>0$.
The six states that are converted to (0,2) states by increasing
$\varepsilon$ show a nearly linear dependence on $\varepsilon$ for $\varepsilon>U$.
Which states develop into a (0,2) state depends on the signs of $\Delta h_S$, $\Delta h_V$, and
$|\Delta h_S|-|\Delta h_V|$, giving rise to $2^3=8$ different configurations to be
distinguished (Table~\ref{table:one}).
\begin{table}[b]
\hfill
\begin{tabular}{|c|>{\columncolor[gray]{.9}}c|c|>{\columncolor[gray]{.9}}c|c|>{\columncolor[gray]{.9}}c|c|c|c|c|}
\hline
  &\tiny $\Delta h_S>$ &\tiny  $\Delta h_V>$ &\tiny   $\Delta h_S>0$ &\tiny  $\Delta h_S>0$&
\tiny $\Delta h_V>0$ &\tiny  $\Delta h_V>0$ &\tiny $0>\Delta h_V$ &\tiny  $0>\Delta h_S$ \\

 &\tiny   $\Delta h_V>0$ &\tiny   $\Delta h_S>0$ &\tiny  $>\Delta h_V$,  &\tiny   $>\Delta h_V$,  &
\tiny  $>\Delta h_S$, &\tiny  $>\Delta h_S$,  &\tiny   $>\Delta h_S$ &\tiny  $>\Delta h_V$  \\

 &   &   &\tiny  $\Delta h_S$  &\tiny  $\Delta h_S$ &
\tiny $\Delta h_V$  &\tiny  $\Delta h_V$   &    &  \\

&   &   &\tiny  $>|\Delta h_V|$  &\tiny  $<|\Delta h_V|$ &
\tiny $>|\Delta h_S|$  &\tiny  $<|\Delta h_S|$   &    &  \\

\hline
\footnotesize
$|\uparrow,\uparrow,+,+\rangle$ &   &   &   &   &   &   &   &  \\
\hline
\footnotesize
$|\uparrow,\uparrow,+,-\rangle$ &   &   & x & x &   &   & x & x\\
\hline
\footnotesize
$|\uparrow,\uparrow,-,+\rangle$ & x & x &   &   & x & x &   &  \\
\hline
\footnotesize
$|\uparrow,\uparrow,-,-\rangle$ &   &   &   &   &   &   &   &  \\
\hline
\footnotesize
$|\uparrow,\downarrow,+,+\rangle$ &   &   &   &   & x & x & x & x\\
\hline
\footnotesize
$|\uparrow,\downarrow,+,-\rangle$ &   &   &   & x &   & x & x & x\\
\hline
\footnotesize
$|\uparrow,\downarrow,-,+\rangle$ &   & x &   &   & x & x & x &  \\
\hline
\footnotesize
$|\uparrow,\downarrow,-,-\rangle$ &   &   &   &   & x & x & x & x\\
\hline
\footnotesize
$|\downarrow,\uparrow,+,+\rangle$ & x & x & x & x &   &   &   &  \\
\hline
\footnotesize
$|\downarrow,\uparrow,+,-\rangle$ & x &   & x & x &   &   &   & x\\
\hline
\footnotesize
$|\downarrow,\uparrow,-,+\rangle$ & x & x & x &   & x &   &   &  \\
\hline
\footnotesize
$|\downarrow,\uparrow,-,-\rangle$ & x & x & x & x &   &   &   &  \\
\hline
\footnotesize
$|\downarrow,\downarrow,+,+\rangle$ &   &   &   &   &   &   &   &  \\
\hline
\footnotesize
$|\downarrow,\downarrow,+,-\rangle$ &   &   & x & x &   &   & x & x\\
\hline
\footnotesize
$|\downarrow,\downarrow,-,+\rangle$ & x & x &   &   & x & x &   &  \\
\hline
\footnotesize
$|\downarrow,\downarrow,-,-\rangle$ &   &   &   &   &   &   &   &  \\
\hline
\end{tabular}
\caption{Charge detection in the presence of the valley degeneracy.
For each of the eight different configurations of $\Delta h_S$ and $\Delta h_V$
the letter x in the table indicates the six basis states that
make a transition to a (0,2) state if $\varepsilon$ is adiabatically
changed from 0 to $\varepsilon>U$. The gray columns belong to the configurations
of the magnetic field which are used in the example in the text.}
\label{table:one}
\end{table}
In the following, we explicitly describe two procedures for
implementing a projective measurement onto one specific state.

For the first procedure we additionally presume that the magnetic field can be
changed in order to reach different configurations for the charge measurement as
given in Table~\ref{table:one}.
This means that after the first charge measurement at $\varepsilon>U$,
which projects the state onto a six- or a ten-dimensional subspace,
and subsequently reducing
$\varepsilon$ to zero, it is possible to change the magnetic field, perform a new
adiabatic transition, and make a new measurement of the charge distribution.
We now consider the example of three charge measurements 
with the following three different configurations of the magnetic field:
(i) $\Delta h_S > \Delta h_V > 0$;
(ii) $\Delta h_S > 0 > \Delta h_V$, $\Delta h_S > |\Delta h_V|$;
(iii) $\Delta h_V > 0 > \Delta h_S$, $\Delta h_V > |\Delta h_S|$.
By considering these three cases in Table \ref{table:one}, one finds that
only the state $|\downarrow,\uparrow, -,  +\rangle$ belongs in all three cases
to the six-dimensional subspace corresponding to a measurement of a (0,2) charge
state.
Therefore, the three charge measurements with outcome (0,2) amount to a projection
on the state.

For the second procedure we use a time-independent magnetic field, for example in
the configuration $\Delta h_S > \Delta h_V > 0$.
Instead of changing the magnetic field, we change the state by single-qubit
operations applied when $\varepsilon=0$.
In our example we may apply $e^{i\pi s_{1x}/2}$, flipping the first spin, after
the first charge measurement and $e^{i\pi s_{2x}/2}$, flipping the second spin,
after the second charge measurement.
The state $|\uparrow,\uparrow,-,+\rangle$ is the only state which is mapped after
the first and after the second spin flip to the six-dimensional subspace which corresponds
to (0,2) states after the adiabatic transition, thus measuring three times a (0,2)
charge configuration is again a projection on one specific state.
If single-qubit operations for all qubits are feasible, any
$|s_1,s_2,\tau_1,\tau_2\rangle$ can be mapped to
$|\downarrow ,\uparrow, -, +\rangle$ or $|\uparrow,\uparrow,-,+\rangle$.
Therefore, a projection on any of these sixteen states can be done in
this way.

\section{Quantum register combining spin and valley qubits}
\label{sec:qregister}
So far, we have shown that universal two-qubit gates between spin qubits or between
valley qubits can be implemented.
Now we consider the situation where both, valley and spin, serve as qubits.
Note that using valley qubits in a quantum register requires valley coherence times
which are sufficiently long to allow for quantum error correction.
This a stricter requirement than in the situation where valley operations are only
needed to achieve spin manipulation (see Sec. \ref{sec:cnot}).
If both kinds of qubits are to be combined in the same quantum register it is
necessary to find a two-qubit gate between a spin and a valley qubit.
Here we show that this can be done by using singlet-triplet qubits in spin and
valley space in one double quantum dot.
For these singlet-triplet qubits the exchange interaction leads directly to a
universal two-qubit gate, as explained in Sec.~\ref{sec:singltripl} below.
Then, in Sec.~\ref{sec:combine}, we show how to connect these qubits
to the usual single-electron spin and valley qubits.
\begin{figure}[t]
\begin{center}
\includegraphics[clip,width=0.85\textwidth]{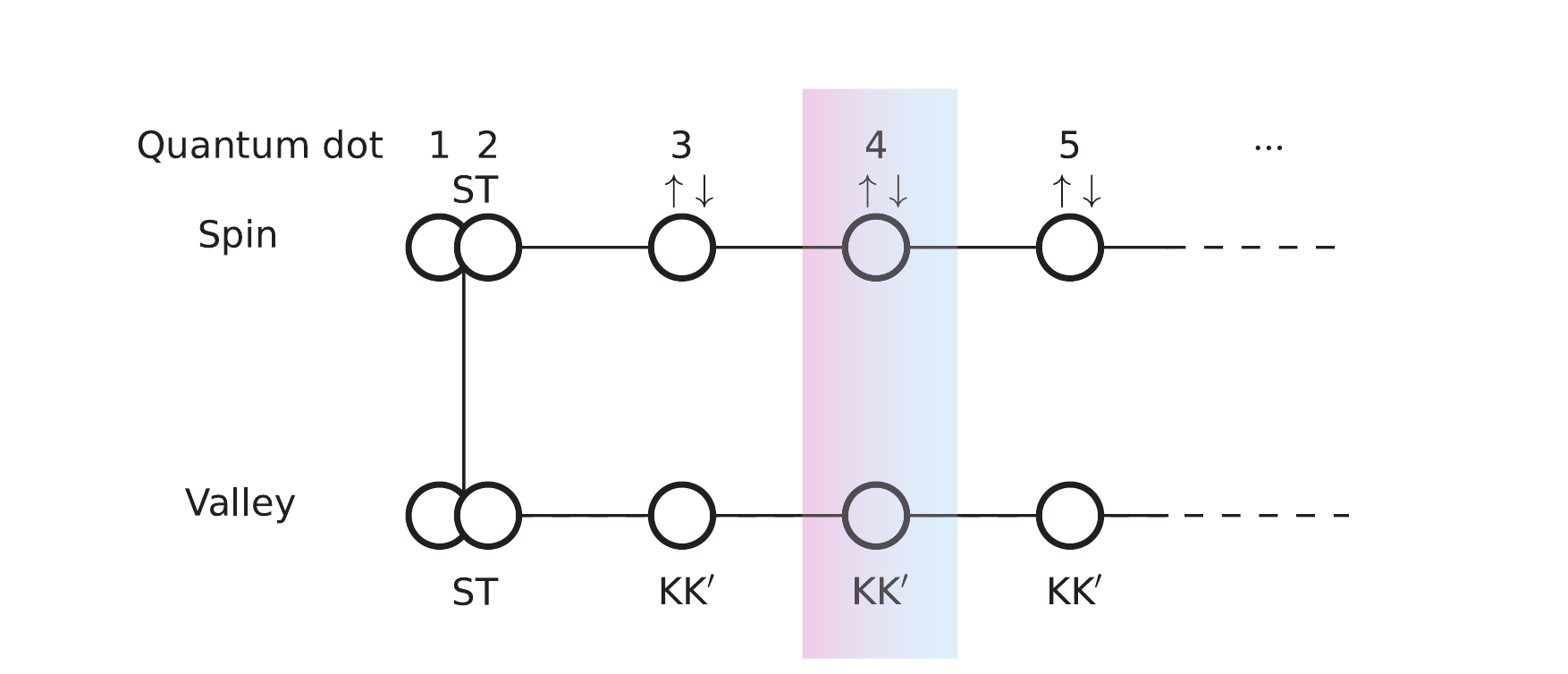}
\caption{Quantum register using both spin and valley qubits.
The shaded area indicates one quantum dot (number 4) occupied with one electron.
Each quantum dot is represented by two circles, one for the spin
and one for the valley iso-spin.
The electrons in quantum dot number 1 and 2 act as singlet-triplet (ST)
qubits to allow universal two-qubit gates between spin and valley.
In all other quantum dots, we consider usual single-electron spin and valley qubits,
$\uparrow\downarrow$ and KK$^\prime$, respectively.
}
\label{fig:architecture}
\end{center}
\end{figure}
This leads effectively to a chain of qubits where nearest neighbors are connected
by universal two-qubit gates, as shown in Fig.~\ref{fig:architecture}.
If $N$ is the number of quantum dots, the number of qubits in this register is
given by $2(N-1)$.

\subsection{Singlet-triplet qubits}
\label{sec:singltripl}
In this subsection, we briefly investigate a different qubit
implementation, in which the singlet state $|\psi_-\rangle\equiv |0\rangle$ and the triplet
state $|\psi_+\rangle \equiv |1\rangle$ (see Eq. (\ref{eqn:bell})) in spin and valley space are
used as the qubit basis states.
Hence, we only consider a subspace of all (1,1) charge states as the logic space.
Since only one out of three triplet states is part of this logic space, the effective
Hamiltonian in the basis
$\{|\psi_-\rangle,|\psi_+\rangle\}_{\rm spin}\otimes\{|\psi_-\rangle,|\psi_+\rangle\}_{\rm valley}$
assumes the simple diagonal form $H_{\rm eff}={\rm diag} (0,-J,-J, 0)$.
Using the Makhlin invariants \cite{makhlin}, it is now easy to show
that the unitary evolution
$U(\pi/2)={\rm diag}(1,i,i,1)$ generated by this Hamiltonian 
is equivalent to a CNOT gate, i.e. it equals CNOT up to single-qubit operations.
Therefore, in this subspace we are able to connect a spin and a valley
qubit with a universal two-qubit gate by applying the exchange interaction.
We define $\sigma_\beta^{(k)}$ ($\beta=x,y,z$) as the Pauli matrices
in the singlet-triplet basis for spin ($k=1$) and valley ($k=2$).
Single-qubit operations can then be performed as follows.
A magnetic field gradient between the dots acts in the
singlet-triplet basis as a single-qubit rotation $\sigma^{(k)}_x$ as any
difference in the Zeeman splitting between the first and the second spin or
valley correspond to a rotation in the singlet-triplet basis.
The gates $\sigma_z^{(k)}$ can be realized by applying the exchange
interaction and valley or spin rotations as
$\exp(i\theta\sigma_z^{(1)})=e^{i\theta}\tau_{1x}U(-2\theta)\tau_{1x}$
and analogously for $\sigma_z^{(2)}$ by replacing $\tau_{1x}$ with $s_{1x}$.
These single-qubit gates together with the universal two-qubit gate allow universal
quantum computing in this two-qubit space.

\subsection{Two-qubit gate between a single-electron and a singlet-triplet qubit}
\label{sec:combine}
In Sec.~\ref{sec:cnot}, we have shown that any two-qubit gate can be applied
between two neighboring spin or valley qubits.
We now consider three quantum dots where spin and valley in the dots number
1 and 2 are prepared in states which are linear combinations of
$|\psi_+\rangle$ and $|\psi_-\rangle$ whereas the spin and the valley of
the third dot can be in any possible state
(Fig.~\ref{fig:architecture}).
To couple the single-spin qubit in dot 3 to the singlet-triplet
spin qubit in dots 1 and 2, we apply a CPHASE gate between the spins
of the electrons in the third
and the second dot where the spin of the third dot is the control qubit.
The spin state of the first and the second quantum dot remains in the subspace
$\{|\psi_+\rangle,|\psi_-\rangle\}$ after this operation.
As $s_{2z}$ represents a change in the relative phase between spins 1
and 2, thus exchanging singlet and triplet states, it acts as a
$\sigma^{(1)}_x$ gate in the singlet-triplet basis, thus this CPHASE gate
between the spins is a CNOT gate in terms of the qubits if they are defined as a
usual spin up/spin down qubit in the third quantum dot and a singlet-triplet
qubit in the first two dots.
A CNOT gate for the valley can be implemented analogously.
Consequently, any two-qubit gate between a usual single-electron 
and a singlet-triplet qubit can be applied.

\section{Conclusions and Outlook}
\label{sec:concs}
In this paper, we have shown that  in the presence of valley
degeneracy, a CNOT gate on spin qubits in a double quantum
dot can be constructed from a sequence of
single-qubit operations and the exchange interaction.
A CNOT gate on the valley qubits can be generated analogously.
For initialization and measurement, an inhomogeneous external
magnetic field is necessary.
A projection on one specific state can be constructed from three charge measurements either
under different configurations of the magnetic field or by using single-qubit
gates.
We could show that adding one double quantum dot in the
singlet-triplet mode allows for a universal quantum
gate (e.g., CNOT)  between a spin and a valley qubit.
This connection between the spin and the valley qubits in a quantum
register implies that universal quantum computing based on spin and valley qubits stored in
the same quantum dots is possible in principle.
Nevertheless, the realization of coherent manipulation of spin and valley qubits
in carbon materials is certainly a big challenge.
An important precondition would be that the valley
degree of freedom has a sufficiently long coherence time, which is
currently unknown. 
An alternative way to create a spin-valley quantum register may lie in
extending the singlet-triplet architecture
with spin and valley degrees of freedom beyond two qubits, e.g., along
the lines of \cite{Hanson2007,Stepanenko2007} for spin-only qubits.

In this work, we have neglected the influence of the spin-orbit interaction, although
it can have important effects in CNT quantum dots \cite{kuemmeth,jespersen}.
It will be a very interesting task to develop a theory for quantum computing
with full orbital and spin degree of freedom in a regime dominated by
spin-orbit coupling.
Despite the proof-of-principle results provided here, there are, obviously,
some remaining open problems regarding the
construction of quantum gates with the exchange interaction with two
degrees of freedom (spin and valley).
It is presently not clear whether there is a shorter sequence for the
$\sqrt{\rm SWAP}$ gate on spin qubits than Eq.~(\ref{eqn:sswap}).
Also, we did not find a \textit{direct} CNOT (or SWAP) gate, i.e.,
without use of singlet-triplet qubits, applied between one single-electron
spin and one single-electron valley qubit, although the exchange interaction also
couples spins and valleys.
Further efforts
could go into finding simpler or even optimal gate implementation for
a spin-valley qubit register.  
The time-evolution operators acting on a four-qubit Hilbert space
are, if we fix the irrelevant global phase, elements of the special unitary
group $SU(16)$, which is a $16^2-1=255$-dimensional space whereas a unitary 
operations on a two-qubit space lies in $SU(4)$, which has only 15 dimensions,
and its two-qubit part can even be described by three real parameters \cite{makhlin,zhang}.
The sequence for the $\sqrt{\rm SWAP}_{\rm spin}$ gate given in
Eq.~(\ref{eqn:sswap}) follows from Eq.~(\ref{eqn:sswap2}), which 
is relatively easy to find as it is constructed as a product
of unitary operations which are diagonal in the double Bell basis.
We now face the more general task of finding a desired quantum gate for a given
sequence of exchange interactions and single-qubit gates where the pulse
lengths (gate times) are free parameters to be determined. 
This can be attempted numerically by minimizing a scalar function which quantifies
the difference between the desired gate and the gate obtained for a given set
of parameters \cite{uqcwtei}.
If the desired gate is an element of $SU(4)\otimes SU(4)\subset
SU(16)$, e.g.,
a two-qubit gate between one spin and one valley iso-spin, we can quantify the
deviation from this subspace and use the Makhlin invariants to describe only
the two-qubit part in both  $SU(4)$ factors.
This reduces the dimension to 231.
Nonetheless, the search for quantum gates constructed with a four-qubit
interaction and single-qubit operations remains a challenging problem.


\section*{Acknowledgements}
The authors acknowledge financial support from DFG
under the program SPP 1285 ``semiconductor spintronics''
and SFB 767 ``functional nanostructures''.


\begin{appendix}
\section{Effective Hamiltonian in the presence of a magnetic field}
\label{sec:appendix}
 The Hamiltonian (\ref{eqn:hamiltonian}) in the presence of a magnetic field
with the same direction in both dots (see Sec. \ref{sec:mf}) can be written
as a $28\times 28$ matrix consisting of 7 independent $4\times 4$ submatrices,
by using the following basis set:
\begin{eqnarray}
{\rm block~1} & & \{c^\dagger_{1,\uparrow +}c^\dagger_{2,\uparrow +}|0\rangle,
                    c^\dagger_{1,\uparrow -}c^\dagger_{2,\uparrow -}|0\rangle,
                    c^\dagger_{1,\downarrow +}c^\dagger_{2,\downarrow +}|0\rangle,
                    c^\dagger_{1,\downarrow -}c^\dagger_{2,\downarrow -}|0\rangle\},\\
{\rm block~2} & & \{c^\dagger_{1,\uparrow +}c^\dagger_{2,\uparrow -}|0\rangle,
                    c^\dagger_{1,\uparrow -}c^\dagger_{2,\uparrow +}|0\rangle,
                    c^\dagger_{1,\uparrow +}c^\dagger_{1,\uparrow -}|0\rangle,
                    c^\dagger_{2,\uparrow +}c^\dagger_{2,\uparrow -}|0\rangle\},\\
{\rm block~3} & & \{c^\dagger_{1,\downarrow +}c^\dagger_{2,\downarrow -}|0\rangle,
                    c^\dagger_{1,\downarrow -}c^\dagger_{2,\downarrow +}|0\rangle,
                    c^\dagger_{1,\downarrow +}c^\dagger_{1,\downarrow -}|0\rangle,
                    c^\dagger_{2,\downarrow +}c^\dagger_{2,\downarrow -}|0\rangle\},\\
{\rm block~4} & & \{c^\dagger_{1,\uparrow +}c^\dagger_{2,\downarrow +}|0\rangle,
                    c^\dagger_{1,\downarrow +}c^\dagger_{2,\uparrow +}|0\rangle,
                    c^\dagger_{1,\uparrow +}c^\dagger_{1,\downarrow +}|0\rangle,
                    c^\dagger_{2,\uparrow +}c^\dagger_{2,\downarrow +}|0\rangle\},\\
{\rm block~5} & & \{c^\dagger_{1,\uparrow -}c^\dagger_{2,\downarrow -}|0\rangle,
                    c^\dagger_{1,\downarrow -}c^\dagger_{2,\uparrow -}|0\rangle,
                    c^\dagger_{1,\uparrow -}c^\dagger_{1,\downarrow -}|0\rangle,
                    c^\dagger_{2,\uparrow -}c^\dagger_{2,\downarrow -}|0\rangle\},\\
{\rm block~6} & & \{c^\dagger_{1,\uparrow +}c^\dagger_{2,\downarrow -}|0\rangle,
                    c^\dagger_{1,\downarrow -}c^\dagger_{2,\uparrow +}|0\rangle,
                    c^\dagger_{1,\uparrow +}c^\dagger_{1,\downarrow -}|0\rangle,
                    c^\dagger_{2,\uparrow +}c^\dagger_{2,\downarrow -}|0\rangle\},\\
{\rm block~7} & & \{c^\dagger_{1,\downarrow +}c^\dagger_{2,\uparrow -}|0\rangle,
                    c^\dagger_{1,\uparrow -}c^\dagger_{2,\downarrow +}|0\rangle,
                    c^\dagger_{1,\downarrow +}c^\dagger_{1,\uparrow -}|0\rangle,
                    c^\dagger_{2,\downarrow +}c^\dagger_{2,\uparrow -}|0\rangle\}.
\end{eqnarray}
We call the 7 submatrices $H_1,\ldots,H_7$ and find
\begin{equation}
 H_1={\rm diag}(2(h_S+h_V),2(h_S-h_V),2(h_V-h_S),-2(h_S+h_V)),
\end{equation}
which is not affected by the exchange interaction as block 1 only contains triplet states, and
\begin{equation}
\label{eqn:matrix}
 H_j = C\mathbbm{1}_4 + \left(\begin{array}{cccc}A &  0 &  t  &  t\\
                                                 0 & -A & -t  & -t\\
                                                 t & -t & U+B &  0\\
                                                 t & -t & 0   & U-B
                              \end{array}
                        \right),
\end{equation}
with
\begin{eqnarray}
 j=2 & : & A=\Delta h_V, B=\varepsilon + \Delta h_S, C=2h_S;\nonumber\\
 j=3 & : & A=\Delta h_V, B=\varepsilon - \Delta h_S, C=-2h_S;\nonumber\\
 j=4 & : & A=\Delta h_S, B=\varepsilon + \Delta h_V, C=2h_V;\nonumber\\
 j=5 & : & A=\Delta h_S, B=\varepsilon - \Delta h_V, C=-2h_V;\nonumber\\
 j=6 & : & A=\Delta h_V + \Delta h_S, B=\varepsilon, C=0;\nonumber\\
 j=7 & : & A=\Delta h_V - \Delta h_S, B=\varepsilon, C=0.\nonumber
\end{eqnarray}
A simple unitary transformation,
\begin{equation}
 W=\frac{1}{\sqrt{2}}\left(\begin{array}{cccc}
         1 & 1  & 0 & 0\\
         1 & -1 & 0 & 0\\
         0 & 0  & 1 & 1\\
         0 & 0 & 1 & -1\\
         \end{array}
\right),
\end{equation}
leads to the matrix form
\begin{equation}
 W H_j W^\dagger =\left(\begin{array}{cccc}
                         0 & A & 0  & 0\\
                         A & 0 & 2t & 0\\
                         0 & 2t & U & B\\
                         0 & 0  & B & U
                        \end{array}
\right)+C\mathbbm{1}_4\hspace{.5cm}(j=2,\ldots,7),
\end{equation}
where the $2\times2$ block in the upper left corner affects only (1,1) while
those in the lower right corner affect only (0,2) charge states.
These blocks are the part $H^{(j)}_0+H^{(j)}_B$ of the Hamiltonian (\ref{eqn:hamiltonian})
where $j=2,\ldots,7$.
The rest of the Hamiltonian, $H_T^{(j)}$, which describes the hopping, couples the subspaces
which are symmetric and asymmetric in charge.
Hamiltonians written in such a matrix form occur already in the spin-only case and have been
considered in Ref. \cite{burkard_imamoglu}, where a Schrieffer-Wolff transformation is used to
derive an effective Hamiltonian for the 16 (1,1) states.
When we omit the index $j$ for better readability, the Schrieffer-Wolff transformation
can be written as $\tilde H = e^{-S}WHW^\dagger e^S\approx H_0+H_B+[H_T,S]/2$ with $S=-S^\dagger$
and $[H_0+H_B,S]=-H_T$.
The approximation holds for $|t|\ll|U\pm B|$.
We can use the result from Ref. \cite{burkard_imamoglu}, and find that
$\tilde H$ is in lowest order given by
two independent $2\times 2$ matrices. 
The matrix describing the subspace with nearly (1,1) charge
distribution has the form
\begin{equation}
 \left(\begin{array}{cc}0 & \tilde A\\
                        \tilde A & -\tilde J\\
       \end{array}
\right)+C\mathbbm{1}_2,
\end{equation}
with
\begin{equation}
 \tilde J=\frac{4t^2U(U^2-B^2-A^2)}{U^4+B^4+A^4-2U^2B^2-2U^2B^2-2B^2A^2},
\end{equation}
and
\begin{equation}
 \tilde A = A\left(1-\frac{J(U^2+B^2-A^2)}{4U(U^2-B^2-A^2)}\right).
\end{equation}
In the case of small gradients in the magnetic field and thus small differences in the Zeeman
splitting between the dots, we can expand these terms and find in lowest order of $\Delta h_V$
and $\Delta h_S$ (the index refers again to the blocks in the basis set)
\begin{equation}
 \tilde J_{2/3}\approx \frac{4t^2U}{U^2-\varepsilon^2}\pm\frac{8t^2U\varepsilon}{(U^2- \varepsilon^2)^2}\Delta h_S
                 = J\left(1\pm \frac{2\varepsilon\Delta h_S}{U^2-\varepsilon^2}\right),
\end{equation}
\begin{equation}
 \tilde J_{4/5}\approx \frac{4t^2U}{U^2-\varepsilon^2}\pm\frac{8t^2U\varepsilon}{(U^2- \varepsilon^2)^2}\Delta h_V
                 = J\left(1\pm \frac{2\varepsilon\Delta h_V}{U^2-\varepsilon^2}\right),
\end{equation}
\begin{equation}
 \tilde J_{6/7}\approx \frac{4t^2U}{U^2-\varepsilon^2} = J,
\end{equation}
and $\tilde A\approx A$.
Expressed with the Pauli matrices for spin and valley this gives the effective Hamiltonian
of Eq. (\ref{eqn:HeffB}) for the states which have approximately (1,1) charge distribution.

Note that we did not use the double Bell basis (see Sec. \ref{sec:xi}) in this Appendix as
this does not provide the matrix form of the Hamiltonian with independent $4\times4$
submatrices in the presence of a magnetic field.
Without a magnetic field the result for the splitting due to exchange interaction is
$\tilde J=4t^2U/(U^2-\varepsilon^2)=J$ for all blocks $2,\ldots,7$.
In this case, the double Bell basis can be obtained by linear combinations of the basis vectors used
here only within the degenerate six- and ten-dimensional subspaces.
Therefore, the effective Hamiltonian is diagonal in the double Bell basis if no magnetic
field is applied.
\end{appendix}

\section*{References} 

\bibliography{spin-valley_QC}{}

\begin{thebibliography}{10}

\bibitem{LoDi1998}
D.~Loss and D.~P. DiVincenzo.
\newblock Quantum computation with quantum dots.
\newblock {\em Phys. Rev. A}, 57:120--126, 1998.

\bibitem{petta}
J.~R. Petta, A.~C. Johnson, J.~M. Taylor, E.~A. Laird, A.~Yacoby, M.~D. Lukin,
  C.~M. Marcus, M.~P. Hanson, and A.~C. Gossard.
\newblock Coherent manipulation of coupled electron spins in semiconductor
  quantum dots.
\newblock {\em Science}, 309(5744):2180--2184, 2005.

\bibitem{koppens}
F.~H.~L. Koppens, C.~Buizert, K.~J. Tielrooij, I.~T. Vink, K.~C. Nowack,
  T.~Meunier, L.~P. Kouwenhoven, and L.~M.~K. Vandersypen.
\newblock Driven coherent oscillations of a single electron spin in a quantum
  dot.
\newblock {\em Nature}, 442:766, 2006.

\bibitem{foletti2009}
S.~Foletti, H.~Bluhm, D.~Mahalu, V.~Umansky, and A.~Yacoby.
\newblock {Universal quantum control of two-electron spin quantum bits using
  dynamic nuclear polarization}.
\newblock {\em Nature Physics}, 5:903--908, 2009.

\bibitem{brunner}
R.~Brunner, Y.-S. Shin, T.~Obata, M.~Pioro-Ladri\`ere, T.~Kubo, K.~Yoshida,
  T.~Taniyama, Y.~Tokura, and S.~Tarucha.
\newblock Two-qubit gate of combined single-spin rotation and interdot spin
  exchange in a double quantum dot.
\newblock {\em Phys. Rev. Lett.}, 107:146801, 2011.

\bibitem{hanson_review}
R.~Hanson, L.~P. Kouwenhoven, J.~R. Petta, S.~Tarucha, and L.~M.~K.
  Vandersypen.
\newblock Spins in few-electron quantum dots.
\newblock {\em Rev. Mod. Phys.}, 79:1217--1265, 2007.

\bibitem{recher_trauzettel}
P.~Recher and B.~Trauzettel.
\newblock Quantum dots and spin qubits in graphene.
\newblock {\em Nanotechnology}, 21(30):302001, 2010.

\bibitem{trauzettel}
B.~Trauzettel, D.~V. Bulaev, D.~Loss, and G.~Burkard.
\newblock Spin qubits in graphene quantum dots.
\newblock {\em Nature Phys.}, 3:192, 2007.

\bibitem{PhysRevLett.105.116801}
J.~G\"uttinger, T.~Frey, C.~Stampfer, T.~Ihn, and K.~Ensslin.
\newblock Spin states in graphene quantum dots.
\newblock {\em Phys. Rev. Lett.}, 105:116801, 2010.

\bibitem{Buitelaar2002}
M.~R. Buitelaar, A.~Bachtold, T.~Nussbaumer, M.~Iqbal, and
  C.~Sch{\"o}nenberger.
\newblock {Multiwall carbon nanotubes as quantum dots}.
\newblock {\em Physical Review Letters}, 88:158801, 2002.

\bibitem{liang}
W.~Liang, M.~Bockrath, and H.~Park.
\newblock Shell filling and exchange coupling in metallic single-walled carbon
  nanotubes.
\newblock {\em Phys. Rev. Lett.}, 88:126801, 2002.

\bibitem{cobden}
D.~H. Cobden and J.~Nyg\aa{}rd.
\newblock Shell filling in closed single-wall carbon nanotube quantum dots.
\newblock {\em Phys. Rev. Lett.}, 89:046803, 2002.

\bibitem{churchill}
H.~O.~H. Churchill, F.~Kuemmeth, J.~W. Harlow, A.~J. Bestwick, E.~I. Rashba,
  K.~Flensberg, C.~H. Stwertka, T.~Taychatanapat, S.~K. Watson, and C.~M.
  Marcus.
\newblock Relaxation and dephasing in a two-electron $^{13}\mathbf{C}$ nanotube
  double quantum dot.
\newblock {\em Phys. Rev. Lett.}, 102:166802, 2009.

\bibitem{palyi2010}
A.~P\'alyi and G.~Burkard.
\newblock Spin-valley blockade in carbon nanotube double quantum dots.
\newblock {\em Phys. Rev. B}, 82:155424, 2009.

\bibitem{reynoso}
A.~A. Reynoso and K.~Flensberg.
\newblock Dephasing and hyperfine interaction in carbon nanotube double quantum
  dots: The clean limit.
\newblock {\em Phys. Rev. B}, 84:205449, 2011.

\bibitem{reynoso2}
A.~A. {Reynoso} and K.~{Flensberg}.
\newblock {Dephasing and Hyperfine Interaction in Carbon Nanotubes Double
  Quantum Dots: The Disordered Case}.
\newblock {\em preprint}, 2012.

\bibitem{schaeffler}
F.~Sch\"affler.
\newblock High-mobility si and ge structures.
\newblock {\em Semiconductor Science and Technology}, 12(12):1515, 1997.

\bibitem{culcer}
D.~Culcer, \L{}. Cywi\ifmmode~\acute{n}\else \'{n}\fi{}ski, Q.~Li, X.~Hu, and
  S.~Das~Sarma.
\newblock Realizing singlet-triplet qubits in multivalley si quantum dots.
\newblock {\em Phys. Rev. B}, 80:205302, 2009.

\bibitem{maue}
B.~M. Maune, M.~G. Borselli, B.~Huang, T.~D. Ladd, P.~W. Deelman, K.~S.
  Holabird, A.~A. Kiselev, I.~Alvarado-Rodriguez, R.~S. Ross, A.~E. Schmitz,
  M.~Sokolich, C.~A. Watson, M.~F. Gyure, and A.~T. Hunter.
\newblock Coherent singlet-triplet oscillations in a silicon-based double
  quantum dot.
\newblock {\em Nature}, 481:344, 2012.

\bibitem{goswami}
S.~Goswami, K.~A. Slinker, M.~Friesen, L.~M. McGuire, J.~L. Truitt, C.~Tahan,
  L.~J. Klein, J.~O. Chu, P.~M. Mooney, D.~W. van~der Weide, R.~Joynt, S.~N.
  Coppersmith, and M.~A. Eriksson.
\newblock Controllable valley splitting in silicon quantum devices.
\newblock {\em Nature Phys.}, 3:41, 2007.

\bibitem{eriksson}
M.~A. Eriksson, M.~Friesen, S.~N. Coppersmith, R.~Joynt, L.~J. Klein,
  K.~Slinker, C.~Tahan, P.~M. Mooney, J.~O. Chu, and S.~J. Koester.
\newblock Spin-based quantum dot quantum computing in silicon.
\newblock {\em Quantum Information Processing}, 3:133--146, 2004.

\bibitem{rycerz}
A.~Rycerz, J.~Tworzydlo, and C.~W.~J. Beenakker.
\newblock Valley filter and valley valve in graphene.
\newblock {\em Nature Phys.}, 3:172, 2007.

\bibitem{PhysRevB.76.235404}
P.~Recher, B.~Trauzettel, A.~Rycerz, Ya.~M. Blanter, C.~W.~J. Beenakker, and
  A.~F. Morpurgo.
\newblock Aharonov-bohm effect and broken valley degeneracy in graphene rings.
\newblock {\em Phys. Rev. B}, 76:235404, 2007.

\bibitem{PhysRevB.79.085407}
P.~Recher, J.~Nilsson, G.~Burkard, and B.~Trauzettel.
\newblock Bound states and magnetic field induced valley splitting in
  gate-tunable graphene quantum dots.
\newblock {\em Phys. Rev. B}, 79:085407, 2009.

\bibitem{PhysRevB.84.195463}
G.~Y. Wu, N.-Y. Lue, and L.~Chang.
\newblock Graphene quantum dots for valley-based quantum computing: A
  feasibility study.
\newblock {\em Phys. Rev. B}, 84:195463, 2011.

\bibitem{burkard_imamoglu}
G.~Burkard and A.~Imamoglu.
\newblock Ultra-long-distance interaction between spin qubits.
\newblock {\em Phys. Rev. B}, 74:041307, 2006.

\bibitem{choi}
M.-S. Choi, R.~L\'opez, and R.~Aguado.
\newblock Su(4) kondo effect in carbon nanotubes.
\newblock {\em Phys. Rev. Lett.}, 95:067204, 2005.

\bibitem{kuemmeth}
F.~Kuemmeth, S.~Ilani, D.~C. Ralph, and P.~L. McEuen.
\newblock Coupling of spin and orbital motion of electrons in carbon nanotubes.
\newblock {\em Nature}, 452:449, 2008.

\bibitem{divincenzo1995}
D.~P. DiVincenzo.
\newblock {Two-bit gates are universal for quantum computation}.
\newblock {\em Physical Review A}, 51(2):1015--1022, 1995.

\bibitem{barenco1995}
A.~Barenco, C.~H. Bennett, R.~Cleve, D.~P. DiVincenzo, N.~Margolus, P.~Shor,
  T.~Sleator, J.~A. Smolin, and H.~Weinfurter.
\newblock {Elementary gates for quantum computation.}
\newblock {\em Physical Review A}, 52(5):3457--3467, 1995.

\bibitem{PhysRevLett.106.086801}
A.~P\'alyi and G.~Burkard.
\newblock Disorder-mediated electron valley resonance in carbon nanotube
  quantum dots.
\newblock {\em Phys. Rev. Lett.}, 106:086801, 2011.

\bibitem{makhlin}
Yu. Makhlin.
\newblock Nonlocal properties of two-qubit gates and mixed states, and the
  optimization of quantum computations.
\newblock {\em Quantum Information Processing}, 1:243--252, 2002.

\bibitem{Hanson2007}
R.~Hanson and G.~Burkard.
\newblock {Universal set of quantum gates for double-dot spin qubits with fixed
  interdot coupling}.
\newblock {\em Physical Review Letters}, 98:050502, 2007.

\bibitem{Stepanenko2007}
D.~Stepanenko and G.~Burkard.
\newblock {Quantum gates between capacitively coupled double quantum dot
  two-spin qubits}.
\newblock {\em Physical Review B}, 75(8):085324, 2007.

\bibitem{jespersen}
T.~S. Jespersen, K.~Grove-Rasmussen, J.~Paaske, K.~Muraki, T.~Fujisawa,
  J.~Nygard, and K.~Flensberg.
\newblock Gate-dependent spin-orbit coupling in multielectron carbon nanotubes.
\newblock {\em Nature Phys.}, 7:348, 2011.

\bibitem{zhang}
J.~Zhang, J.~Vala, S.~Sastry, and K.~B. Whaley.
\newblock Geometric theory of nonlocal two-qubit operations.
\newblock {\em Phys. Rev. A}, 67:042313, 2003.

\bibitem{uqcwtei}
D.~P. DiVincenzo, D.~Bacon, J.~Kempe, G.~Burkard, and K.~B. Whaley.
\newblock Universal quantum computation with the exchange interaction.
\newblock {\em Nature}, 408:339, 2000.

\end{thebibliography}

\end{document}